\documentclass[preprint]{imsart}
\pdfoutput=1
\usepackage[latin1]{inputenc}
\usepackage[T1]{fontenc}

\usepackage{latexsym,amsmath,amssymb,amscd,amsfonts,amsthm,mathrsfs}
\usepackage{graphics,natbib}
\usepackage{a4wide}
\usepackage{color,geometry}
\usepackage{latexsym}
\usepackage{longtable}
\usepackage{float}
\usepackage{url}
\usepackage{graphicx}
\usepackage{dsfont,multirow,float,bm,lscape}

\def \ybf {\mathbf y}
\def \Vbf {\mathbf V}
\def \pibf {\bm{\pi}}
\DeclareMathOperator*{\diag}{diag}

\begin{document}

\begin{frontmatter}

\title{Comparing Model Selection and Regularization Approaches to Variable Selection in Model-Based Clustering}

\begin{aug}
\author{\fnms{Gilles} \snm{Celeux}\thanksref{t1}\ead[label=e1]{gilles.celeux@math.u-psud.fr}},
\author{\fnms{Marie-Laure} \snm{Martin-Magniette}\thanksref{t2,t3}\ead[label=e2]{marie\_laure.martin@agroparistech.fr}},\\
\author{\fnms{Cathy} \snm{Maugis-Rabusseau}\thanksref{t4}\ead[label=e3]{cathy.maugis@insa-toulouse.fr}}
\and
\author{\fnms{Adrian E.} \snm{Raftery}\thanksref{t5,t6}\ead[label=e4]{raftery@u.washington.edu}\ead[label=u1,url]{http://www.foo.com}}

\thankstext{t1}{Inria Saclay - {\^I}le-de-France, Orsay, France}
\thankstext{t2}{UMR INRA 1165 - UEVE, ERL CNRS 8196, Unit{\'e} de Recherche en G{\'e}nomique V{\'e}g{\'e}tale, Evry, France}
\thankstext{t3}{UMR AgroParisTech/INRA MIA 518, Paris, France}
\thankstext{t4}{Institut de Math{\'e}matiques de Toulouse, INSA de Toulouse, Universit{\'e} de Toulouse}
\thankstext{t5}{Department of Statistics, University of Washington, Seattle, Washington, USA}
\thankstext{t6}{School of Mathematical Sciences, University College Dublin, Ireland.}
\runauthor{G. Celeux et al.}
\end{aug}

\begin{abstract}
We compare two major approaches to variable selection in
clustering: model selection and regularization. Based on previous results,
we select the method of \citet{Maugis&2009b},
which modified the method of \citet{RafteryDean2006},
as a current state of the art model selection method.
We select the method of \citet{WittenTibshirani2010}
as a current state of the art regularization method.
We compared the methods by simulation in terms of their accuracy in
both classification and variable selection. In the first simulation
experiment all the variables were conditionally independent given cluster
membership. We found that variable selection (of either kind) yielded
substantial gains in classification accuracy when the clusters were well
separated, but few gains when the clusters were close together.
We found that the two variable selection methods had comparable
classification accuracy, but that the model selection approach had
substantially better accuracy in selecting variables.
In our second simulation experiment, there were correlations among
the variables given the cluster memberships. We found that the model selection approach was substantially more accurate in terms of both classification and variable selection than the regularization approach, and that both gave more accurate classifications than $K$-means without variable selection.
\end{abstract}

\begin{keyword}
\kwd{Model-based clustering, Model selection, Regularization approach, Variable selection}
\end{keyword}

\begin{keyword}[class=AMS]
\kwd{62H30}
\end{keyword}

\end{frontmatter}

\section{Introduction}
Over the past 20 years, model-based clustering
\citep{Wolfe1970,McLachlanBasford1988,BanfieldRaftery1993,CeleuxGovaert1995,FraleyRaftery2002}
has come to rival the heuristic clustering methods
such as single link, complete link and $K$-means that dominated previously.
In the past decade it has been realized that the performance of model-based
clustering can be degraded if irrelevant or noise variables are present.
As a result, there has been considerable interest in variable selection
for model-based clustering.

Two of the most used general approaches have been model selection and
regularization. Model selection approaches to the problem were pioneered
by \citet{Law&2004} and \citet{Tadesse&2005}, who partitioned the
set of candidate variables into two sets, one set relevant to the clustering
and the other irrelevant. They assumed that the irrelevant variables
were statistically independent of the relevant ones.

\citet{RafteryDean2006} --- hereafter RD ---
realized that irrelevant variables are often
correlated with relevant ones, and developed a model selection method
that takes account of this, and a greedy search algorithm to implement it.
Their method assumes that each irrelevant variable
depends on all the relevant variables according to a linear regression model.

\citet{Maugis&2009a} pointed out
that the Raftery-Dean method implies a very non-parsimonious model
for all the variables jointly, explaining the method's lacklustre
performance in some comparative studies
\citep{SteinleyBrusco2008,WittenTibshirani2010}.
They proposed modifying it by
selecting the predictor variables in the linear regression part of the model.
\citet{Maugis&2009b} went further and allowed explicitly for an irrelevant
variable to be independent of all the relevant variables.
Although at first sight these may not seem like major changes to the method,
they can actually make a big difference to the results
and have led to greatly improved performance.
The resulting method provides both more parsimonious and realistic models.
Following \citet{Celeux&2011}, we refer to it here as the RD-MCM method.

A different approach, via regularization, was proposed by
\citet{PanShen2007} \citep[see also][]{Xie&2008,WangZhu2008,Zhou&2009,Guo&2010}.
Another regularization approach, called sparse $K$-means (SparseKmeans),
was proposed by \citet{WittenTibshirani2010} ---
hereafter WT; this can be viewed as a simpler version
of the Clustering on Subsets of Objects (COSA) approach of
\citet{FriedmanMeulman2004}. WT found in a
simulation study that their method outperformed both COSA and
the regularization method of \citet{PanShen2007}.

WT also found that their method outperformed the RD method.
However, this appears to have been due
to the nonparsimonious model underlying the RD method,
and the WT method gave similar results to the RD-MCM method
under the simulation setup of WT \citep{Celeux&2011}.

Overall, it seems that, among variable selection methods for model-based
clustering, the RD-MCM method is currently one of the best performing   
model selection methods, and the WT method is one of the best performing regularization methods.  
In this paper, we compare these two methods in a range of simulated
and real data settings.

In Section \ref{sect-methods} we summarize the two methods we are comparing.
In Section \ref{sect-simulation} we give results for a range of simulation
setups previously proposed, and in
Section \ref{sect-realdata} we give results for a waveform dataset and
for data from a genetics experiment.
In Section \ref{sect-discussion} we discuss
limitations, caveats, other approaches  and possible extensions of our results.

\section{Variable Selection Methods for Model-Based Clustering}
\label{sect-methods}

\subsection{Model Selection Methods}
\label{sect-methods.modsel}

Let $n$ observations $\ybf=(y_1,\ldots,y_n)$ to be clustered be described
by $p$ continuous variables ($y_i\in\mathbb{R}^p$).
In the model-based clustering framework, the multivariate continuous data
$\ybf$ are assumed to come from several subpopulations (clusters) modeled with a multivariate
Gaussian density. The observations are assumed to arise from a finite Gaussian mixture with $K$ components and a model $m$,  
namely
$$
    f(y_i|K,m,\alpha)=\underset{k=1}{\stackrel{K}{\sum}} \pi_k \phi(y_i|\mu_k,\Sigma_{k(m)}),
$$
where $\pibf=(\pi_1,\ldots,\pi_K)$ is the mixing proportion vector ($\pi_k\in(0,1)$ for all
$k=1,\ldots,K$ and $\sum_{k=1}^K \pi_k=1$), $\phi(.|\mu_k,\Sigma_k)$ is the $p$-dimensional Gaussian
density function with mean $\mu_k$ and variance matrix $\Sigma_k$, and
$\alpha=(\pibf,\mu_1,\ldots,\mu_K,\Sigma_1,\ldots,\Sigma_K)$ is
the parameter vector.

This framework yields a range of possible models $m$, each
corresponding to different assumptions on the forms of the
covariance matrices, arising from a modified spectral decomposition.
These include whether the volume, shape and orientation of each mixture
component vary between components or are constant across clusters
\citep{BanfieldRaftery1993,CeleuxGovaert1995}.
Typically, the mixture parameters are estimated via maximum likelihood
using the EM algorithm, and both the model structure and the number
of components $K$ are chosen using the BIC or other penalized likelihood
criteria \citep{FraleyRaftery2002}.
Software to implement this methodology includes the
{\sc Mclust} R package (\url{http://www.stat.washington.edu/mclust/}),
and the {\sc mixmod} software (\url{http://www.mixmod.org}).
The latter implements 28 Gaussian mixture models, most of which
are also available in {\sc Mclust}.
Here we view each mixture component as corresponding to one cluster,
and so the term cluster is used hereafter.

The RD-MCM method, as described by \citet{Maugis&2009b},
involves three possible roles for the variables:
the relevant clustering variables ($S$), the
redundant variables ($U$) and the independent variables ($W$).
Moreover, the redundant variables $U$ are explained by a subset $R$
of the relevant variables $S$, while the variables $W$ are assumed
to be independent of the relevant variables.
Thus the data density is assumed to be decomposed into three parts as follows:
$$
    f(y_i|K,m,r,l,\Vbf,\theta) = \underset{k=1}{\stackrel{K}{\sum}} \pi_k \phi(y_i^S|\mu_k,\Sigma_{k(m)}) \times \phi(y_i^U|a+y_i^R b,\Omega_{(r)}) \times \phi(y_i^W|\gamma,\tau_{(\ell)})
$$
where $\theta=(\alpha,a,b,\Omega,\gamma,\tau)$ is the full parameter vector
and $\Vbf = (S,R,U,W)$.
We denote the form of the regression variance matrix $\Omega$ by $r$;
it can be spherical, diagonal or general.
The form of the variance matrix $\tau$ of the independent variables $W$
is denoted by $\ell$ and can be spherical or diagonal.

The RD-MCM method recasts
the variable selection problem for model-based clustering as a
model selection problem. This model selection problem is
solved using a model selection criterion, decomposed into the sum of the
three values of the BIC criterion associated with the Gaussian mixture,
the linear regression and the independent Gaussian density respectively.
The method is implemented by using two backward stepwise algorithms for
variable selection, one each for the clustering and the linear regression.
A backward algorithm allows one to start with all variables in order to
take variable dependencies into account. A forward procedure, starting
with an empty clustering variable set or a small variable subset, could be
preferred for numerical reasons if the number of variables is large.
The method is implemented in the {\it SelvarClustIndep} software.\footnote{{\it SelvarClustIndep} is available at \url{http://www.math.univ-toulouse.fr/~maugis/}}

The RD-MCM method generalizes several previous model selection methods.
The procedure of \citet{Law&2004}, where irrelevant
variables are assumed to be independent of all the relevant variables, corresponds to $W=S^c$, $R=\emptyset$,
$U=\emptyset$.   The RD method \citep{RafteryDean2006}
assumes that the irrelevant variables are
regressed on the whole relevant variable set ($W=\emptyset$, $U=S^c$ and $R=S$).
The generalization of \citet{Maugis&2009a}
enriches this model by allowing the irrelevant variables to be explained by
only a subset of the relevant variables $R \subset S$ $(W=\emptyset, U=S^c)$;
this method is implemented in the {\it  SelvarClust}
software\footnote{{\it  SelvarClust} is available at \url{http://www.math.univ-toulouse.fr/~maugis/}}.

\subsection{Regularization Methods}
\label{sect-methods.regularization}
A review of sparse clustering techniques can be found in WT.
Most of these methods embed sparse clustering in the model-based
clustering framework.  Notable exceptions are the COSA approach
of \citet{FriedmanMeulman2004} and the WT approach,
which can be thought of as a simpler version
of the approach of \citet{FriedmanMeulman2004}.

WT propose a sparse clustering procedure, called the {\it Sparse $K$-means}
algorithm. This procedure is based
on a variable weighting in the $K$-means algorithm. Let $\mathcal{P}=(\mathcal{P}_1,\ldots,\mathcal{P}_K)$ denote a
clustering of observations and $n_k$ the number of observations in cluster $\mathcal{P}_k$. The {\it sparse $K$-means} algorithm maximizes a weighted between-cluster sum of squares
$$
\underset{\mathcal{P},\mathbf{w}}{\max} \underset{j=1}{\stackrel{p}{\sum}} w_j \left(\frac 1 n \underset{i=1}{\stackrel{n}{\sum}}\underset{i'=1}{\stackrel{n}{\sum}} (y_{ij}-y_{i'j})^2   - \underset{k=1}{\stackrel{K}{\sum}} \frac{1}{n_k} \underset{i,i'\in\mathcal{P}_k}{\sum} (y_{ij}-y_{i'j})^2\right) ,
$$
where $\mathbf{w}=(w_1,\ldots,w_p)$ is the weight vector such that for all $j$, $w_j\geq 0$, $\|\mathbf{w}\|^2 \leq 1$ and $\|\mathbf{w}\|_1\leq t$ where $t$ is a tuning parameter. This parameter is chosen by a permutation approach
using the gap statistic of \citet{Tibshirani&2001}.
Their method is implemented in the R package {\sc sparcl}.

\section{Simulation Experiments}\label{sect-simulation}

We now give comparative results for two simulation experiments
with setups based on simulation experiments in the related literature.
We compare three methods: $K$-means with no variable selection,
SparseKmeans, and the RD-MCM method.

\subsection{Simulation Experiment 1: Conditionally Independent Variables}
This simulated example is inspired by WT's Simulation 2.
It concerns the situation where the relevant variables are conditionally
independent given the cluster memberships, and the irrelevant
variables are independent both of the relevant variables and
of one another.

Five scenarios are considered with $p=25$ variables in Scenarios 1-4 and
$p=100$ variables in Scenario 5.
The first five variables are distributed according to a mixture of three
equiprobable spherical Gaussian distributions $\mathcal{N}(\mu_k,I_{5})$
with $\mu_1=-\mu_2=(\mu,\ldots,\mu)\in\mathbb{R}^5$ and $\mu_3=0_5$.
Twenty (respectively ninety-five) noisy standard centered Gaussian variables are appended in Scenarios 1-4 (respectively in Scenario 5).

The number of observations is $n=30$
in Scenarios 1 and 2, and $n=300$ in Scenarios 3, 4 and 5.
In Scenarios 1 and 3, $\mu=0.6$, while $\mu=1.7$ in Scenarios 2, 4 and 5.
Note that the second scenario is the one
considered by WT. Since the SparseKmeans method requires the user
to know the number of clusters, the true number $K=3$
is assumed known for all three methods.
Moreover, the true mixture shape is fixed for the RD-MCM method.
Twenty-five datasets were simulated for each scenario.

We evaluate the three methods according to three criteria.
Classification performance is evaluated by the Adjusted Rand Index (ARI)
\citep{HubertArabie1985}, reported in percent.
The ARI is 100 for a perfect classification and
0 for a random one; larger ARI is better.
Performance in selecting the right variables is evaluated using the
Variable Selection Error Rate (VSER), defined as the average number
of errors in selecting (or not selecting) variables, as a percentage of the
total number of variables considered.
SparseKmeans is defined as selecting a variable if the corresponding
weight is positive. Smaller values of VSER are better. Finally, we also report the average
number of variables selected (\#VarSel). The true number of variables in this
experiment is 5, so the closer \#VarSel is to 5 the better.

The results are shown in Table \ref{tbl-sim1}.
For classification, both variable selection methods greatly outperformed
$K$-means without
variable selection for Scenarios 2 and 4 (with $\mu=1.7$),
while for Scenarios 1 and 3 (with $\mu=0.6$) there was not a big difference.
The two variable selection methods performed similarly
for classification across the first four scenarios.
In Scenario 5, the three methods had similar behaviour.

For variable selection, however, there were clear distinctions.
The RD-MCM method had a better variable selection error rate than
SparseKmeans in the five scenarios, with a substantial difference
for the larger sample size (Scenario 5).
The number of variables selected by RD-MCM was closer to the
true number (5) for all five scenarios. SparseKmeans selected
substantially more variables in Scenarios 1-3 and all the variables
in Scenarios 4-5.

\begin{figure}[htbp]
\centerline{\includegraphics[width=\textwidth]{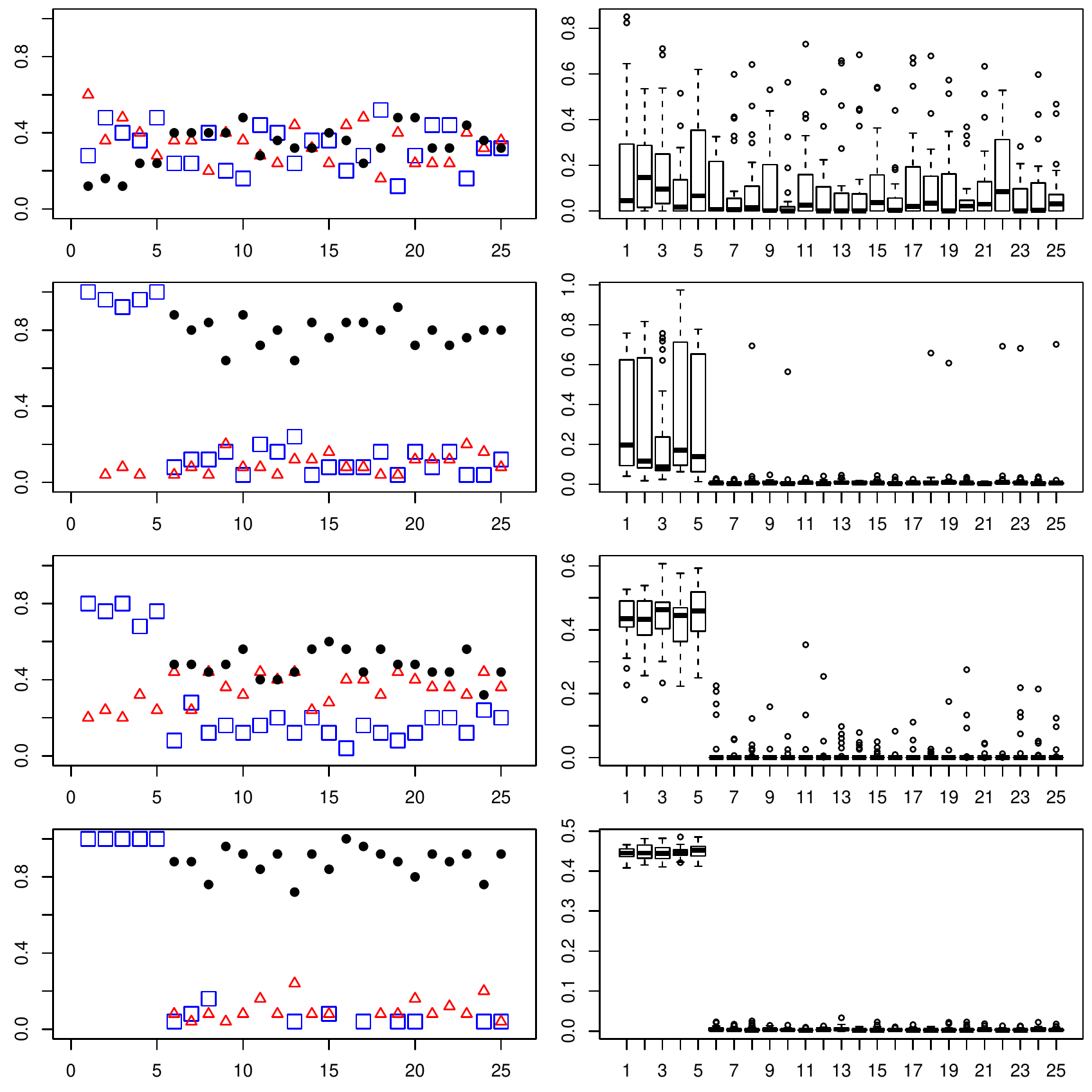}}
\caption{Simulation Experiment 1: On the left, the proportion of times each variable was declared relevant (square), redundant (triangle) or independent
(circle) by RD-MCM in the first four scenarios
(Scenario 1 at top to Scenario 4 at bottom).
Zero values are not shown.
On the right, boxplots of the weights given by SparseKmeans for each variable in the four scenarios.}
\label{FigSim1-1}
\end{figure}

Figure \ref{FigSim1-1} displays the distribution of the variable roles
with RD-MCM and the variations of the weights given by SparseKmeans
for Scenarios 1-4. In Scenario 1, neither method accurately distinguished
between clustering and non-clustering variables, which is not surprising
as  both the sample size and the between-cluster separation were small.
In Scenario 2 the RD-MCM method identified the five clustering variables
correctly in most cases, while the weights from SparseKmeans varied
considerably. In Scenarios 3 and 4, both methods distinguished clearly
between clustering and non-clustering methods, with RD-MCM selecting them
and not the non-clustering variables, and SparseKmeans giving much
higher weights on average to the clustering than the non-clustering variables.
SparseKmeans gave positive, although small, weights to non-clustering
variables in many cases.

\begin{figure}[htbp]
\centerline{\includegraphics[width=\textwidth]{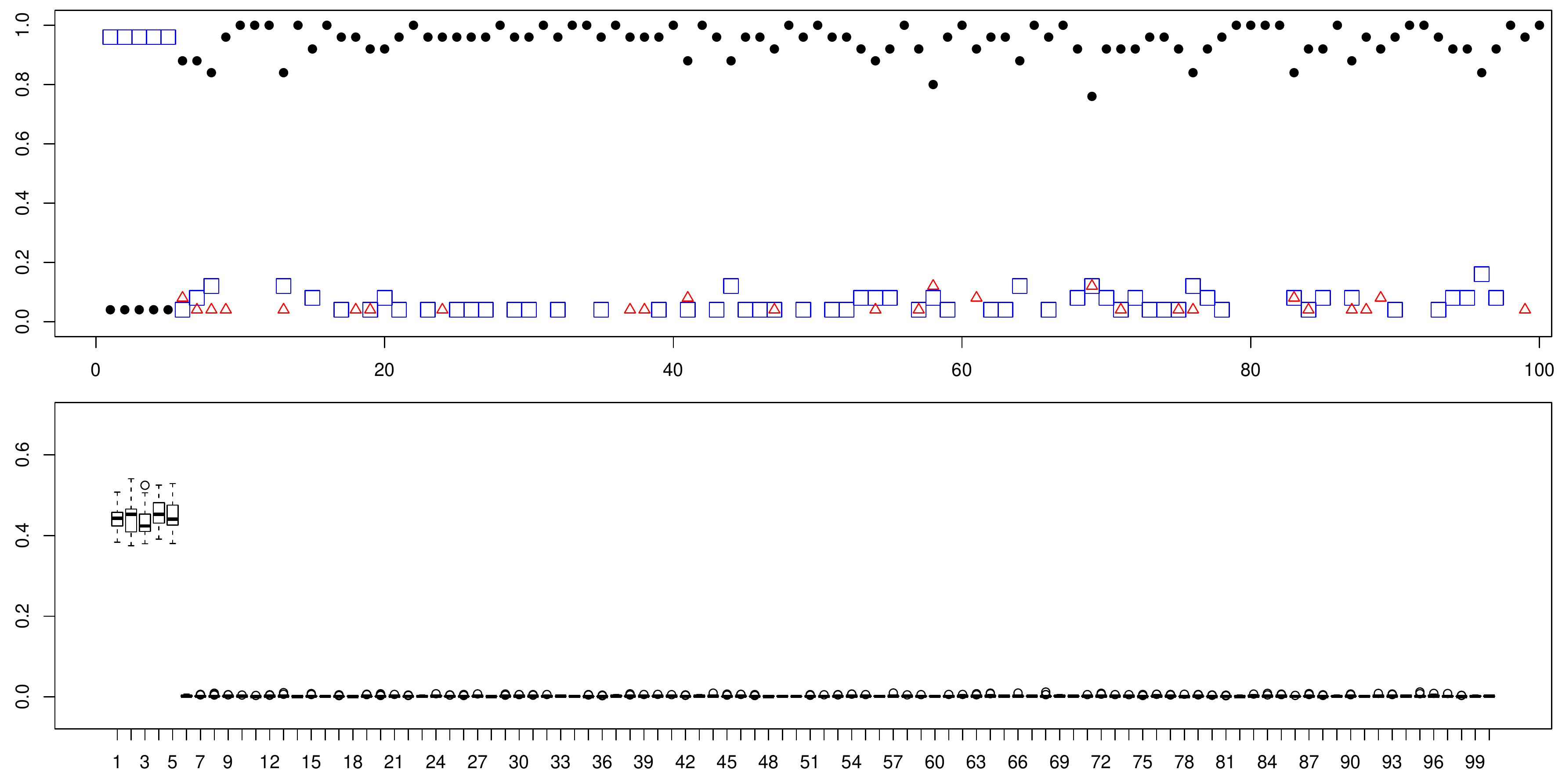}}
\caption{Simulation Experiment 1: On the top, proportion of times each
variable was declared  relevant (square), redundant (triangle) or independent
(circle) by RD-MCM in Scenario 5. Zero values are not shown.
On the bottom, boxplots of the weights given by SparseKmeans
for each variable in Scenario 5.}
\label{FigSim1-2}
\end{figure}

Figure \ref{FigSim1-2} shows the results for Scenario 5, with 100 variables,
of which five were relevant. RD-MCM selected the correct variables in
most cases, and also correctly identified the non-clustering variables
as independent (rather than just redundant). SparseKmeans gave
much higher weights to the clustering variables than to the non-clustering
ones on average, although again the weights for the non-clustering
variables were always positive.

\begin{table}[htbp]
\begin{center}
\caption{\label{tbl-sim1}
Simulation Experiment 1 Results. ARI is the Adjusted Rand Index
expressed in percent (the higher the better),
VSER is the variable selection error rate in percent (the lower the better).
\#VarSel is the average number of variables selected
(correct number $=5$).
Standard errors are shown in parentheses.
The method performing best in each scenario under each criterion
is shown in bold.}
\begin{tabular}{ccclccc} \\ \hline
Scenario    & $n$ &$\mu$& Method                                  & ARI & VSER & \#VarSel \\ \hline
1           & 30  & 0.6 & $K$-means                                  & {\bf 11 (9)} & 80 (---)     & 25.0 (---) \\
25 variables&     &     & SparseKmeans                            & 8 (7)        & 45 (22)      & 14.4 (6.3) \\
            &     &     & RD-MCM                                  & 8 (7)        & {\bf 36 (11)}& {\bf 8.1 (1.9)} \\
\hline
2            & 30  & 1.7 & $K$-means                                 & 44 (10)      & 80 (---)     & 25.0 (---) \\
25 variables &     &     & SparseKmeans                           & {\bf 81 (17)}& {\bf 13 (16)}& 8.2 (4.0) \\
             &     &     & RD-MCM                                 & 71 (24)      & 17 (12)      & {\bf 6.8 (1.4)} \\
\hline
3            & 300 & 0.6 & $K$-means                                 & 20 (3)       & 80 (---)     & 25.0 (---) \\
25 variables &     &     & SparseKmeans                           & 14 (4)       & 76 (11)      & 24.0 (2.7) \\
             &     &     & RD-MCM                                 & {\bf 23 (3)} & {\bf 10 (7)} & {\bf 7.0 (1.8)} \\
\hline
4            & 300 & 1.7 & $K$-means                                 & 64 (14)       & 80 (---)   & 25.0 (---) \\
25 variables &     &     & SparseKmeans                           & {\bf 89 (3)}  & 80 (0)     & 25.0 (0) \\
             &     &     & RD-MCM                                 & 88 (3)        & {\bf 2 (3)} & {\bf 5.6 (0.9)} \\
\hline
5            & 300 & 1.7 & $K$-means                                 & 86 (3)       & 95 (---)    & 100.0 (---) \\
100 variables&     &     & SparseKmeans                           & {\bf 88 (4)} & 95 (0)      & 100.0 (0) \\
             &     &     & RD-MCM                                 & 84 (18)      & {\bf 4 (2)} & {\bf 8.4 (2.3)} \\
\hline
\end{tabular}
\end{center}
\end{table}

\subsection{Simulation Experiment 2: Correlated Variables}
We now consider situations where the variables are correlated
conditionally on the cluster memberships.
Two of the seven simulated situations considered in
\cite[Sect. 6.1]{Maugis&2009b} are now considered.
The $n=2000$ observations are described by $p=14$ variables.
The first two variables are distributed according to a mixture of four Gaussian distributions
$\mathcal{N}(\mu_{k},I_2)$ with $\mu_{1}=(0,0)$, $\mu_{2}=(4,0)$,
$\mu_{3}=(0,2)$ and $\mu_4=(4,2)$.
In the first situation \cite[Scenario 3]{Maugis&2009b},
the mixing proportion vector is $\pibf=(0.2,0.3,0.3,0.2)$.
The last twelve variables are simulated as follows:
$$
\left\{
\begin{array}{l}
y_i^{3}= 3 y_i^1 + \varepsilon_i \textrm{ with } \varepsilon_i\sim\mathcal{N}(0,0.5), \\
y_i^{\{4-14\}}\sim \mathcal{N}((0,0.4,0.8,\ldots,3.6,4),I_{11}).
\end{array}
\right.
$$

In the second situation, an equiprobable mixture is considered on the first two variables and the last twelve variables are simulated as follows:
\begin{equation}
\left\{
\begin{array}{l}
y_i^{\{3-11\}}= (0,0,0.4,0.8,\ldots,2) + y_i^{\{1,2\}}\, b + \varepsilon_i \textrm{ with } \varepsilon_i\sim\mathcal{N}(0_9,\Omega), \\
y_i^{\{12-14\}}\sim \mathcal{N}((3.2,3.6,4),I_{3}),
\end{array}
\right.
\label{eq-simex2.2}
\end{equation}
where the regression coefficients are
$$b=((0.5,1)',(2,0)',(0,3)',(-1,2)',(2,-4)',(0.5,0)',(4,0.5)',(3,0)',(2,1)').$$
In (\ref{eq-simex2.2}), $\Omega=\diag(I_3,0.5I_2,\Omega_1,\Omega_2)$ is the block diagonal regression variance matrix with
$\Omega_1=\mbox{Rot}(\pi/3)'\diag(1,3)\mbox{Rot}(\pi/3)$
and $\Omega_2=\mbox{Rot}(\pi/6)'\diag(2,6)\mbox{Rot}(\pi/6)$,
where $\mbox{Rot}(\theta)$ denotes the plane rotation matrix with angle
$\theta$.

The third situation is analogous to the second situation with many more noisy variables:
$$
\left\{
\begin{array}{l}
y_i^{\{3-11\}}= (0,0,0.4,0.8,\ldots,2) + y_i^{\{1,2\}}\, b + \varepsilon_i \textrm{ with } \varepsilon_i\sim\mathcal{N}(0_9,\Omega)\\
y_i^{\{12-41\}} \sim\mathcal{N}(0_{30},I_{30}); y_i^{\{42-71\}} \sim\mathcal{N}(2_{30},I_{30}); y_i^{\{72-101\}} \sim\mathcal{N}(4_{30},I_{30})
\end{array}
\right.
$$

\begin{figure}[htbp]
\centerline{\includegraphics[width=0.7\textwidth]{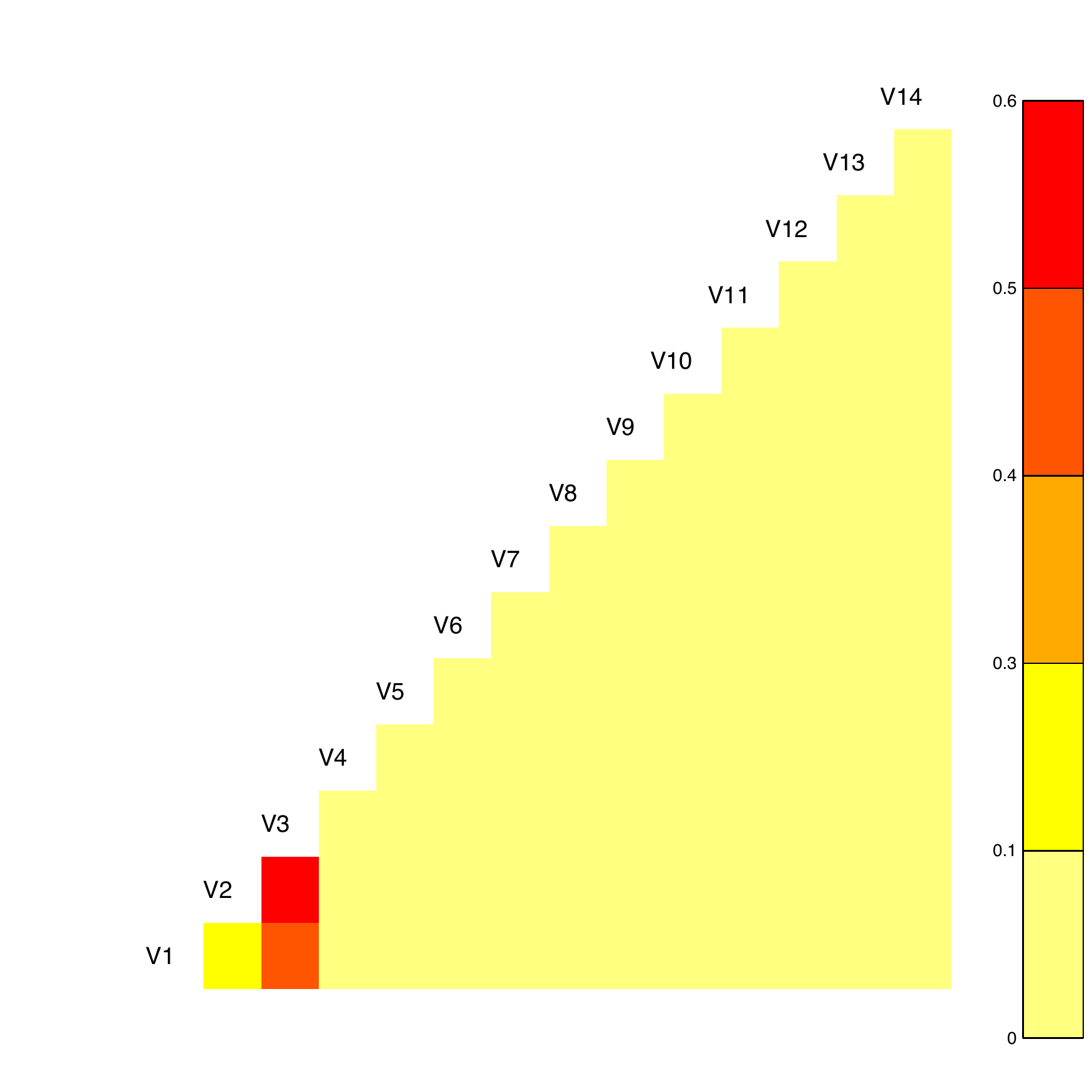}}
\vspace*{-0.5cm}
\centerline{\includegraphics[width=0.6\textwidth]{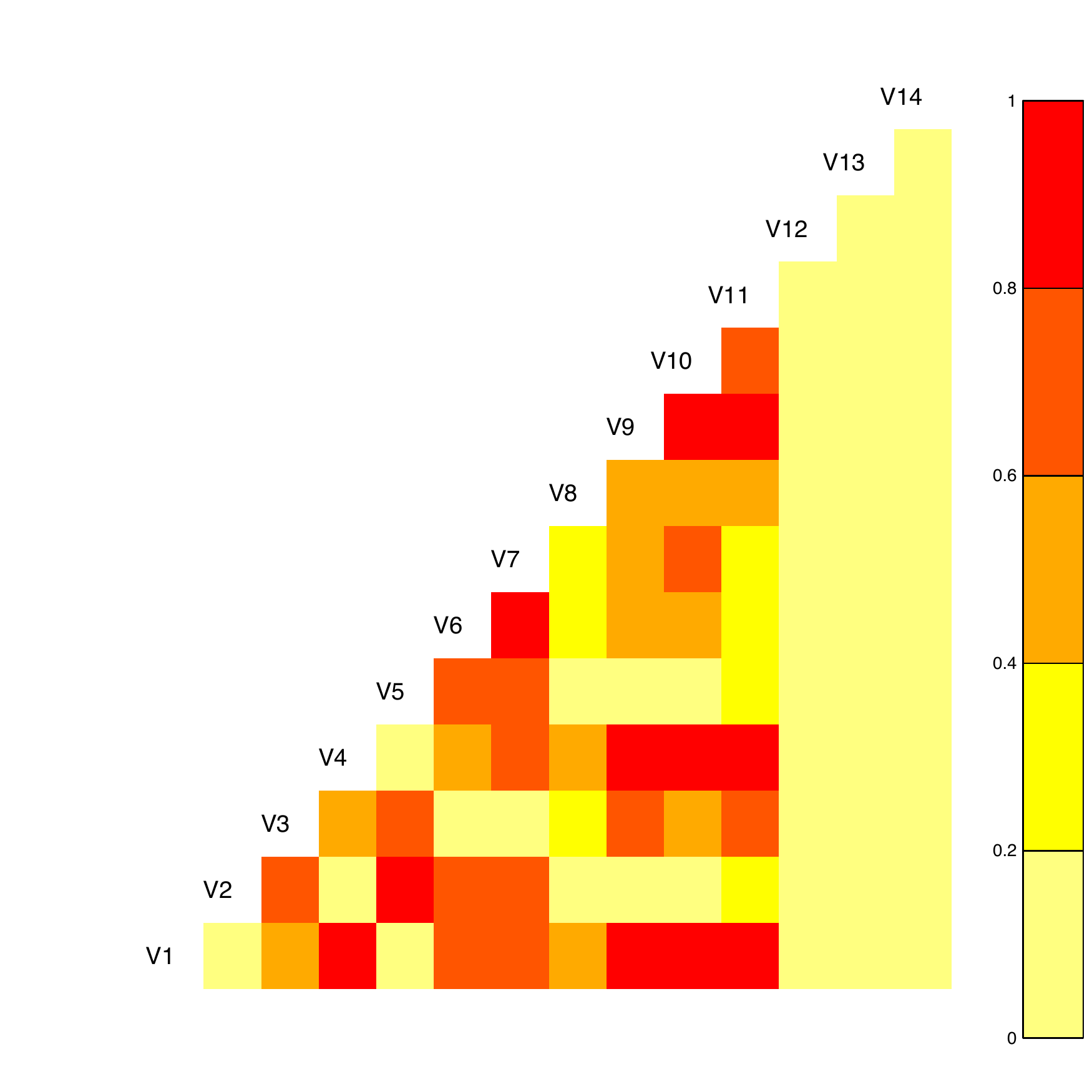}}
\caption{Simulation Experiment 2: Correlation matrix for all 14 variables
in one simulation of Scenario 1 (top) and Scenarios 2 and 3 (bottom).}
\label{CorrPlot-Ex2-2}
\end{figure}

The correlations of the variables in the three situations are shown in
Figure \ref{CorrPlot-Ex2-2}. The true number of clusters, $K=4$, is assumed
known for all the procedures. Each situation has been replicated 50 times.
The results are shown in Table \ref{tbl-sim2}.

For both classification and variable selection, RD-MCM substantially
outperformed both $K$-means without variable selection, and
SparseKmeans.
This may be because the simulation experiment involves
correlation between the variables beyond the clustering.
Neither $K$-means nor SparseKmeans takes account of this explicitly,
while RD-MCM does.

\begin{figure}[htbp]
\centerline{\includegraphics[width=\textwidth]{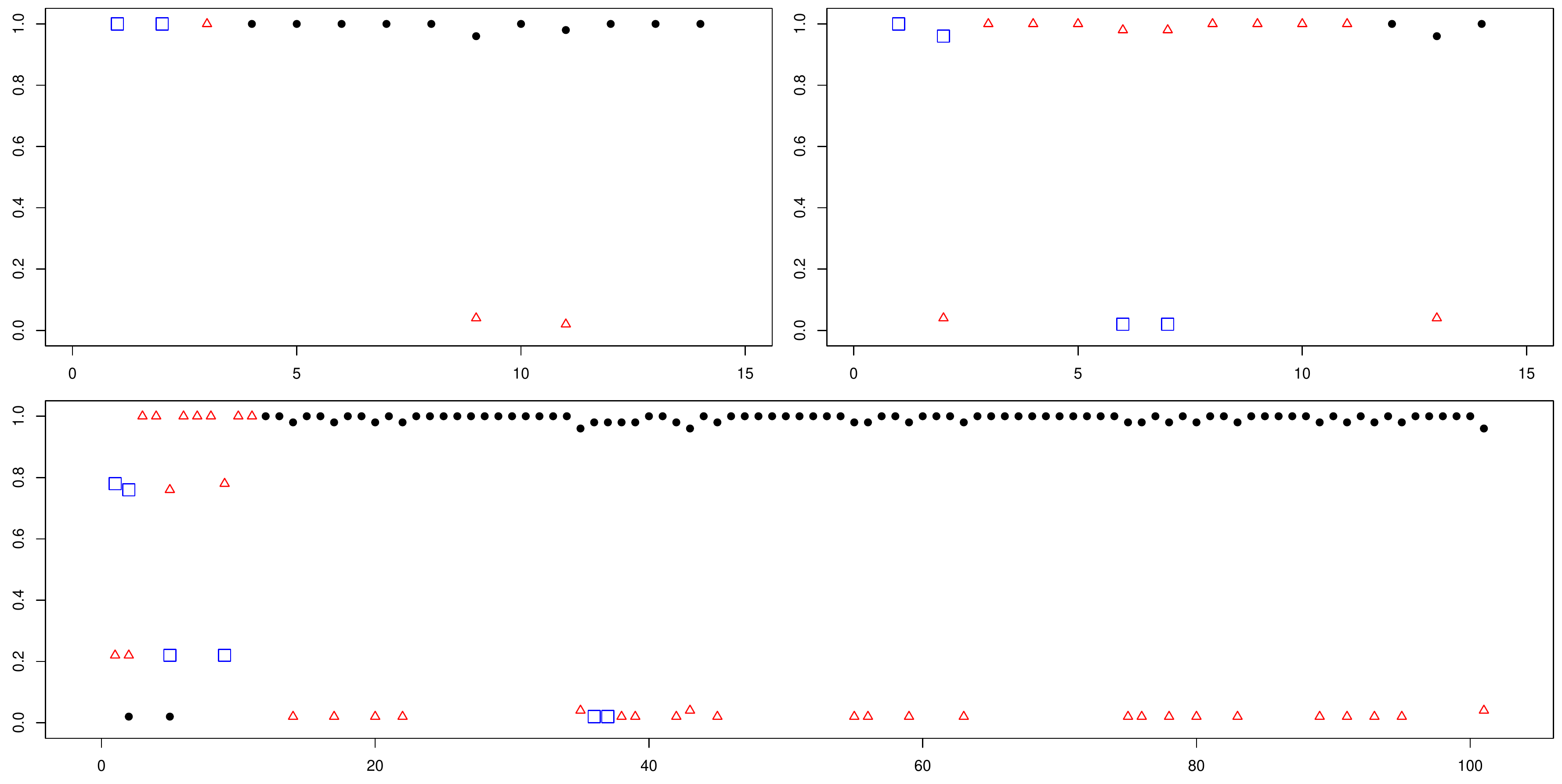}}
\caption{Simulation Experiment 2: Proportion of times each variable was declared  relevant (blue square), redundant (red triangle) or independent (black circle) by RD-MCM in Scenario 1 (top left), Scenario 2 (top right) and Scenario 3 (bottom). Zero values
are not shown. In each scenario, only the first two variables were relevant.
}
 \label{FigSim2-3}
\end{figure}

\begin{figure}[htbp]
\centerline{\includegraphics[width=\textwidth]{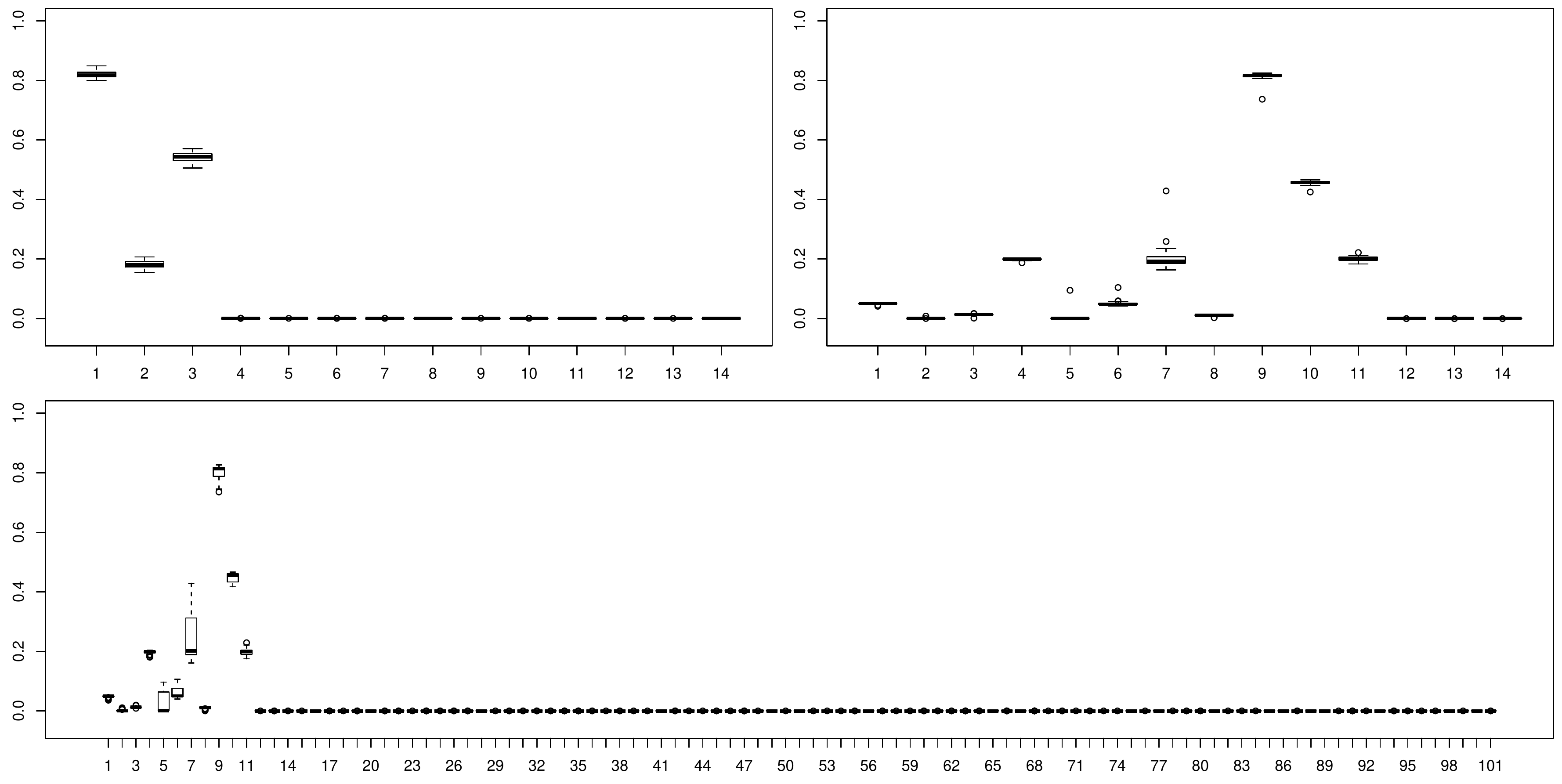}}
\caption{Simulation Experiment 2: Boxplots of the weights $\mathbf{w}$ given by SparseKmeans for each variable in Scenario 1 (top left), Scenario 2 (top right) and Scenario 3 (bottom) respectively. In each scenario, only the first two variables were relevant.}
 \label{FigSim2-1}
\end{figure}

Figures \ref{FigSim2-3} and \ref{FigSim2-1} display the proportion of times
each variable was declared relevant, redundant or independent by RD-MCM,
and the variations of the weights given by SparseKmeans.
In most cases, RD-MCM correctly selected the clustering variables
and not the non-clustering variables in all three scenarios.
SparseKmeans tended to give high weights to redundant variables,
and in Scenarios 2 and 3 it gave low weights on average
to the first two variables, which are the relevant ones.

\begin{table}[htbp]
\begin{center}
\caption{\label{tbl-sim2}
Simulation Experiment 2 Results. ARI is the Adjusted Rand Index
in percent (the higher the better),
VSER is the variable selection error rate in percent (the lower the better).
\#VarSel is the average number of variables selected
(correct number $=2$).
Standard errors are into parentheses.
The method performing best in each scenario under each criterion
is shown in bold.}
\begin{tabular}{clccc} \\ \hline
Scenario & Method                             & ARI          & VSER        & \#VarSel \\
\hline
1 & $K$-means                                    & 52 (1)       & 86 (---)    & 14.0 (---) \\
  & SparseKmeans     & 47 (2)                 & 86 (0)      & 14.0 (0) \\
  & RD-MCM                                    & {\bf 57 (4)} & {\bf 0 (0)} & {\bf 2 (0)} \\
\hline
2 & $K$-means                                    & 57 (2)       & 86 (---)    & 14.0 (---) \\
  & SparseKmeans                              & 31 (3)       & 85 (5)      & 13.8 (1) \\
  & RD-MCM                                    & {\bf 60 (2)} & {\bf 1 (3)} & {\bf 2.0 (0.2)} \\
\hline
3 & $K$-means                                    & 56 (2)       & 86 (---)    & 101.0 (---) \\
  & SparseKmeans $(w>0)$                      & 34 (6)       & 80 (36)      & 82.9 (36.5) \\
  & RD-MCM                                    & {\bf 57 (7)} & {\bf 1 (2)} & {\bf 2.02 (0.14)} \\
\hline
\end{tabular}
\end{center}
\end{table}

\section{Real Data Examples}
\label{sect-realdata}

\subsection{Waveform data}
\label{sect-realdata.waveform}
This dataset consists of $n=5,000$ points with $p=40$ variables.
The first 21 variables are relevant to the classification problem,
and are based on a random convex
combination of two of three waveforms
sampled at the integers $\{1,\ldots,21\}$
with noise added. Nineteen noisy standard centered Gaussian
variables are appended. A detailed description of the waveform
dataset is given by \citet[][pp.43--49]{Breiman1984}.

To compare the RD-MCM method with SparseKmeans on an equal footing,
we assume that the correct number of clusters, $K=3$, is known.
The RD-MCM method selects clustering variables
$\hat{S} = \{4-7, 9-18\}$, redundant variables
$\hat{U} = \{2, 3, 8, 19, 20, 38\}$, and independent variables
$\hat{W} = \{1, 21-37, 39, 40\}$. For the SparseKmeans, all variables are selected because all weights are positive.
The Adjusted Rand Index between the RD-MCM clustering and
the three waves is 0.25, while for SparseKmeans it is 0.23.
Thus RD-MCM performs slightly better in terms of classification,
while using fewer variables..

Note that the RD-MCM method can select not only the clustering
variables, but also the number of clusters, if this is unknown.
This is unlike SparseKmeans, which requires the number of clusters
to be assumed or known in advance.
If we do not assume the number of clusters known here,
the RD-MCM method selects $K=6$ clusters with clustering variables
$\hat S=\{4-18\}$, redundant variables
$\hat U=\{2,3,19,20,38\}$ that are explained by the predictor variables
$\hat R=\{5-7, 9-12, 14,15,17 \}$,
and independent variables $\hat W=\{1,21-37, 39,40 \}$.

We compare these results with the SparseKmeans method where $K=6$ is fixed.
In this case also, SparseKmeans selects all the variables.
The ARI between the RD-MCM six-cluster solution and the three waves is 0.257,
which is actually slightly better than with the three-cluster solution;
and is 0.27 for SparseKmeans, also better than when the number of clusters is
taken to be three. This suggests that in some cases there may be some
advantage to selecting the number of clusters based on the data,
even when the true number of clusters is known a priori.

\subsection{Transcriptome data}

\label{sect-realdata.transcriptome}
We turn to a transcriptome dataset of {\it Arabidopsis thaliana}, extracted from the
database CATdb \citep{Gagnot&2008} which has been considered by
\cite{Maugis&2009c} for clustering.
This dataset consists of $n=4616$ genes described by $p=33$ biotic stress experiments (partitioned into $9$ biological projects).
Each gene is described by a vector $y_i\in\mathbb{R}^{33}$,
where $y_{ij}$ is the test statistic from
the experiment $j$ for the differential analysis.
For details of the normalization and the differential
analysis steps, see \citet{Gagnot&2008}.

We first analyzed these data using the RD-MCM method.
Based on previous analyses of transcriptome datasets \citep{Maugis&2009a},
we allowed the number of mixture components to vary between 10 and 30,
and we considered the mixture forms EEE (identical variance matrices) and VEE
(different volumes and the same orientation and shape for the variance
matrices) in the notation of \citet{FraleyRaftery2002}.
The method selected the VEE model with $\hat{K}=26$ components.
Experiments $22$, $23$ and $26$, which come from the same biological project,
were identified as redundant and the other 30 experiments
were declared relevant for clustering.
The three redundant experiments were explained by the relevant variables
$\hat{R}=\{1, 5, 8, 13, 24, 25, 27, 28-31, 33\}$.
They were mainly explained by experiments $24$, $25$ and $27$,
which come from the same biological project.
The sizes of the 26 clusters varied widely, from 45 to 602 genes per cluster.

\begin{figure}
\centerline{\includegraphics[width=\textwidth]{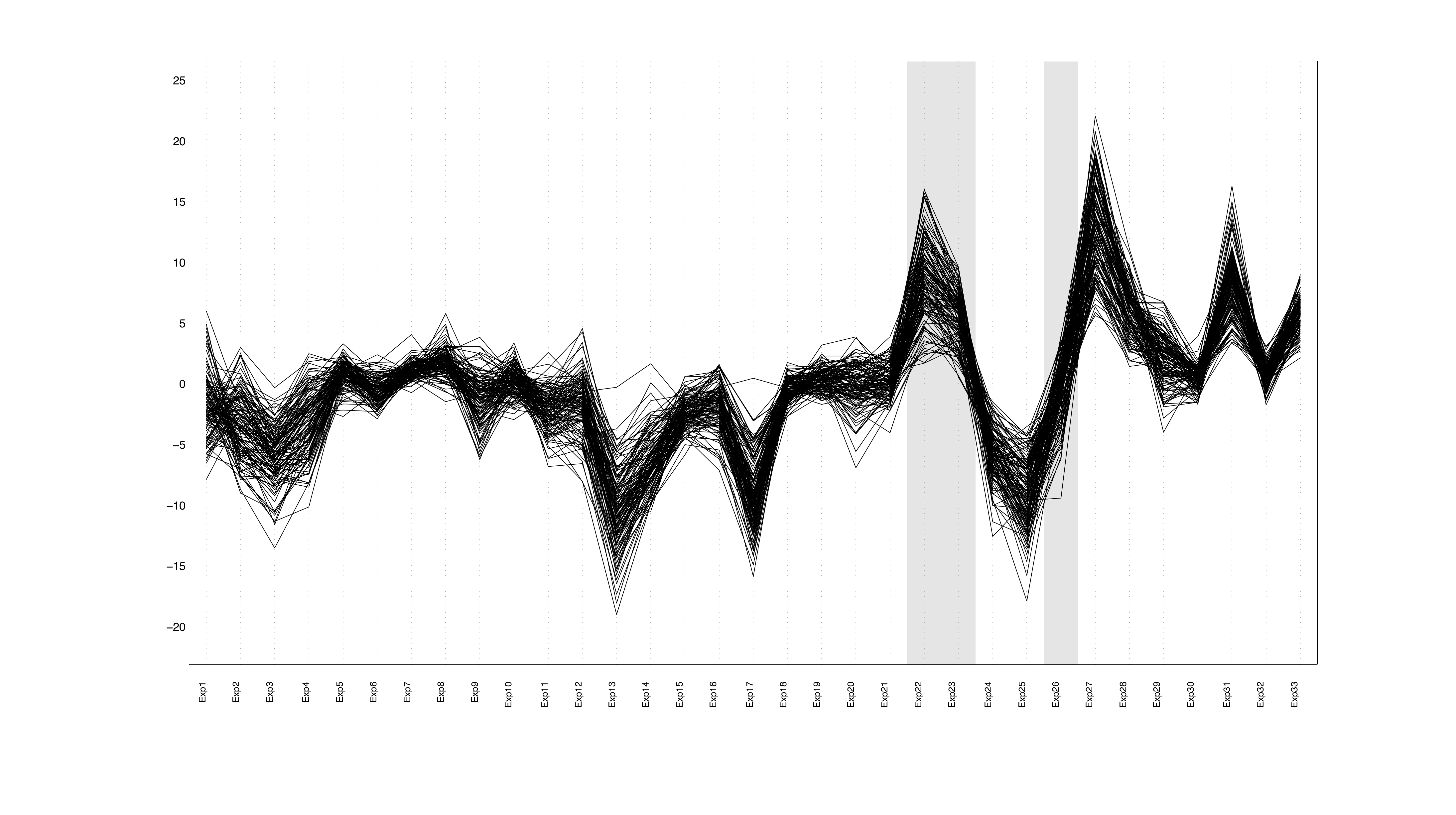}}
\caption{Transcriptome Data: Gene Profiles in Cluster 19}\label{Cluster19}
\end{figure}

As an example, Figure~\ref{Cluster19} shows the expression profiles
of the genes in cluster 19, which are quite homogeneous and
clearly coexpressed. This corresponds to other biological
information, because most of the genes in cluster 19 are in the cell nucleus.
In addition, the biological function of 101 of the 129 genes in cluster 19
is linked to transcription. Of the remaining 28 genes, the role of eight
is unknown according to the functional annotation.
Thus this cluster analysis can potentially shed light on the
biological functions of these eight genes for which this is not currently
known \citep{Maugis&2009c}.

The results from the SparseKmeans and $K$-means procedures were quite
different from one another.
Repeating the SparseKmeans procedure from random initial positions with $K=26$
led to an average ARI of $0.135 (SD\ 0.002)$ compared with the RD-MCM
clustering.
For instance, the 129 genes of cluster 19 of the RD-MCM solution shown in
Figure~\ref{Cluster19} were divided between cluster 8 (100 genes among 193),
cluster 21 (3 genes among 358),
cluster 22 (1 gene among 224) and cluster 26 (25 genes among 320)
in the ``best'' solution provided by SparseKmeans.
It was also surprising to see that the ARI between
SparseKmeans and $K$-means was rather low at 0.349.
On the other hand, the ARI between the RD-MCM partition and the partition
obtained with the VEE model with $\hat{K}=26$ components without selecting
the variables was higher (0.578), suggesting more stable results.

\section{Discussion}
\label{sect-discussion}
We have carried out a comparison between model selection and regularization
approaches to variable selection in model-based clustering.
These are two of the leading approaches.
For each general approach we have selected a specific method based on
results from previous results in the literature.
The model selection method is the method of \citet{Maugis&2009b}
which modified that of \citet{RafteryDean2006}; we refer to it as
the RD-MCM method. The regularization method is the SparseKmeans method of
\citet{WittenTibshirani2010}. We also compared them with $K$-means without
variable selection.

We compared the methods by simulation in terms of their accuracy in
both classification and variable selection. In the first simulation experiment
all the variables were conditionally independent given cluster membership.
We found that variable selection (of either kind) yielded substantial
gains in classification accuracy when the clusters were well separated,
but few gains when the clusters were close together.
We found that the two variable selection methods had comparable
classification accuracy, but that the model selection approach
had substantially better accuracy in selecting variables.

In our second simulation experiment, there were correlations among the
variables given the cluster memberships. We found that the model selection
approach was substantially more accurate in terms of both classification
and variable selection than the regularization approach, and that both
gave more accurate classifications than $K$-means without
variable selection.

Another advantage of the model selection approach is that it allows one
to select the number of clusters based on the data, while the
regularization approach requires that it be known or specified
in advance by the user.
It also allows one to select among a range of models for the covariance
structure. Also, we found that the results of SparseKmeans were quite
sensitive to the tuning parameter.

There are several other recent proposals for variable selection in model-based
clustering.
\citet{Fraiman&2008} proposed a method for variable selection
{\it after} the clustering has been carried out that also assumes
the number of clusters known. Thus it is not fully comparable with the
methods considered here, which carry out clustering and variable selection
simultaneously. \citet{NiaDavison2012} proposed a fully Bayesian approach using
a spike and slab prior, while \citet{Kim&2012} proposed a Bayesian
approach using Bayes factors to compare the different models.
\citet{LeeLi2012} proposed an approach to variable selection for
model-based clustering using ridgelines.

It is also worth mentioning the dimension reduction methods of
\citet{Bouveyron&2007,McNicholasMurphy2008,Scrucca2010,McLachlan&2011} for model-based clustering.
Their goal is not variable selection.
Also, \citet{Poon&2013} proposed a method for facet determination
in model-based clustering; this is related to but not the same as
variable selection.
There are also several other recent proposals for variable selection through regularisation. See for instance
\citet{GMV09,Sun12,BBa13}. Finally, it is worth mentioning the review paper \citet{BBb13}.

\paragraph{Acknowledgements:} Raftery's research was supported by
NIH grants R01 GM084163, R01 HD054511 and R01 HD070936,
as well as by Science Foundation Ireland E.T.S.~Walton Visitor Award
11/W.1/I2079.

\bibliographystyle{imsart-nameyear}
\bibliography{varselcomp20130725ar}
\end{document}